\documentclass[prl,aps,twocolumn]{revtex4-2}
\usepackage{float}
\usepackage{dynkin-diagrams}
\usepackage{hyperref}
\usepackage{amsmath}
\usepackage{amsfonts}
\usepackage{xcolor}
\usepackage{dynkin-diagrams}
\usepackage{amssymb}
\usepackage{graphicx}

\begin{document}

\title{Bosonic M-Theory From a Kac-Moody Algebra Perspective}
\author{Keith Glennon}
\email{keith.glennon@oist.jp}
\affiliation{Okinawa Institute of Science and Technology}
\date{\today}

\begin{abstract}

We study the existence of a `bosonic m-theory' extension of the 10D and 26D closed bosonic string in terms of Kac-Moody algebras. We argue that $K_{11}$ and $K_{27}$ are symmetries which protect the coefficients of the closed bosonic string in 10 and 26 dimensions. Therefore the Susskind-Horowitz `bosonic m-theory' obtained by compactification on $S^1/Z_2$, which does not produce the correct coefficients, must be replaced by something that preserves $K_{11}$ and $K_{27}$. We argue that in 11D, a non-trivial `bosonic m-theory' should be considered as (the bosonic sector) of m-theory, and in 27D that no obvious `bosonic m-theory' exists.
\end{abstract}

\maketitle

\section{1: E Theory Background}

It has been conjectured that m-theory in eleven dimensions possesses the Kac-Moody group $E_{11}$ as a symmetry in the low-energy limit [1]. In [2] it was shown that the bosonic sector of the low-energy effective action (\textbf{LEEA}) of m-theory arises from the non-linear realization of $E_{11} \otimes_s l_1 / I_c(E_{11})$. Here $l_1$ is the vector representation of $E_{11}$, and $I_c(E_{11})$ is a certain Cartan-involution invariant subgroup of $E_{11}$. In this approach, the theory of the 10D IIA and IIB LEEA's immediately arise from $E_{11}$ as different decompositions of the $E_{11}$ algebra [3]. We thus see that the transition from, say, the IIA theory to m-theory, at low energies, from the perspective of $E_{11}$, amounts to a simple re-writing of the underlying $E_{11}$ symmetry group, rather than the usual argument based on the dilaton measuring the scale of the compactified dimension [4]. 
\par
A review of this work is given in [5], an explanation of how to do many of the computations in detail was given in [6], and a more general introduction to this approach is given in [7], sometimes referred to as `E theory'. This approach has been applied to gravity in [8] and the closed bosonic string (\textbf{CBS}) in [9]. Specifically, in [9] it was shown that the dynamics of the LEEA of the CBS in 26D arise from the non-linear realization of $K_{27} \otimes_s l_1 / I_c(K_{27})$. Here $K_{27}$ is the very-extended Kac-Moody group $D_{24}^{+++}$. 
\par
Although the computations in [9] were done in $D=26$ dimensions for concreteness, as noted in the discussion, the results generalize to dimension $D$ for the Kac-Moody group $K_{D+1} = D_{D-2}^{+++}$ and we will assume this in what follows.

\section{2: A Review of `Bosonic M-Theory' via Compactification on $S^1/Z_2$}

The basic idea behind `bosonic m-theory' can be summarized via the following analogy. The action for the 26D CBS possesses a dilaton, as the IIA theory does. The 26D CBS LEEA in the Einstein frame takes a similar form to that of the LEEA of the 10D IIA string, both actions being scaled by a dilaton-dependent factor $e^{-\sigma}$. Based on this, in [10] it was proposed that the 26D CBS can directly be interpreted as the compactification of a to-be-determined 27D `bosonic m-theory' LEEA. This theory was assumed to be based on both gravity and a three-form gauge field, and a 27D action was proposed in [10]. Although the procedure was motivated by m-theory compactified on $S^1$ into IIA, the authors compactified on $S^1/Z_2$ (akin to the heterotic string) to eliminate undesirable terms so that their 27D action to reduced into a 26D action similar to the CBS. 
\par 
This procedure, as inspired as it was, unfortunately did not result in the correct coefficient for the dilaton in the action. It was argued that, due to the lack of a symmetry like supersymmetry ({\it or say any other symmetry principle that would protect the coefficients}), this coefficient may change under renormalization and so its value need not be correct.

\section{3: Arguing Against Compactification on $S^1/Z_2$}

A year later, taking the incorrect coefficient at face value, in [11] it was argued that the action of [10] does not possess any of the coset symmetries of the 26D CBS predicted by the (then conjectured) $K_{27}$ symmetry [1], and that this is due to the incorrect dilaton coefficient. Compare this to the situation between $E_{11}$ and the IIA theory: $E_{11}$ contains the coset symmetries of the 10D IIA theory, of the 11D m-theory LEEA, the dynamics of (the bosonic sectors of) both theories, and it does all this without requiring supersymmetry. $E_{11}$ in fact acts as a replacement for supersymmetry when restricted to the bosonic sector, it predicts the fields and dynamics of the bosonic sector of the LEEA with the correct coefficients and possesses all correct coset symmetries, describing both the 11D m-theory LEEA and the 10D IIA LEEA. We can thus say that $E_{11}$ is a symmetry that protects the values of the coefficients in the 11D m-theory LEEA, and the 10D IIA LEEA. 
\par 
Given the recent claim [9] that $K_{27}$ is a symmetry of the 26D CBS, we can now argue that $K_{27}$ is a symmetry that protects the values of the coefficients of the 26D CBS. Similarly, $K_{11}$ protects the coefficients of the 10D CBS. Therefore, `bosonic m-theory', as interpreted as a compactification on $S^1/Z_2$, simply cannot be the correct approach to a presumptive `bosonic m-theory', as it does not contain $K_{11}$ in 11D nor $K_{27}$ in 27D.
\par 
From this perspective we demand that, whatever `bosonic m-theory' is, if it exists, it is expected to possess a symmetry group that contains the symmetry groups ($K_{11}$ in 11D and $K_{27}$ in 27D) which protect the coefficients of the CBS action and contains the correct coset symmetries. These assumptions are implicit in the relationship between the IIA string and $E_{11}$, in fact even between the IIB string and $E_{11}$ despite the latter not being related by direct compactification, so it is natural to expect such properties. Thus, Kac-Moody algebras can function as a guiding light in seeking an alternative approach to `bosonic m-theory'.

\section{4: A Kac-Moody Approach to Discovering an M-Theory LEEA}

In this paper we will use the tools of Kac-Moody algebras to try to discover a presumptive `bosonic m-theory' by preserving the Kac-Moody symmeteries of the CBS at all times. We will first review how the 10D `IIA Algebra' extends into $E_{11}$ in 11D. We will then look for similar ways to extend $K_{11}$ (underlying the 10D CBS) into an 11D algebra that will describe the presumptive 11D `bosonic m-theory'. We will then try to repeat this for the 26D CBS starting from $K_{27}$. We are going to find that the only non-trivial extension of $K_{11}$ is going to be $E_{11}$, and so `bosonic m-theory' in 11D is (the bosonic sector of) m-theory. In 27D, we will find that $K_{27}$ should be extended into $E_{27}$ yet it is not a subalgebra of $E_{27}$, that a theory of dynamics based on $E_{27}$ cannot exist, and that no other obvious extension of $K_{27}$ exists, so that `bosonic m-theory' in 27D does not seem to exist.
\par 
The basic pattern is as follows. If the dynamics of a $D$-dimensional theory can be described by the nonlinear realization of the form $\hat{G} \otimes_s l_1 / I_c(\hat{G})$, where $\hat{G}$ is a Lorentzian Kac-Moody algebra [7], and $l_1$ is its vector representation, then its m-theory extension in $(D+1)$-dimensions should also be described by a nonlinear realization of the form $\hat{G}' \otimes_s l_1' / I_c(\hat{G}')$, where $\hat{G}'$ at least contains $\hat{G}$. However any old group $\hat{G}'$ containing $\hat{G}$ will not do, we should somehow try to mimic the way that the IIA algebra in 10D extends into $E_{11}$ in 11D. Therefore, we will first review how this is done in the following section, and the reader may refer to [7] for more detail.

\section{5: From the `IIA Algebra' to $E_{11}$}

The Dynkin diagram of $E_{11}$, decomposed with respect to its $A_{10}$ subgroup, is given by
$$
\begin{matrix}
& & &  & & & & & & & & & & & \oplus & 11 & & \\
& & &  & & &  & & & & & & & & | & & & \\
\bullet & - & \bullet & - & \bullet & - & \bullet & - & \bullet & - & \bullet & - & \bullet & - & \bullet & - & \bullet & - & \bullet \\
1 & & 2 & & 3 & & 4 & & 5 & & 6 & & 7 & & 8 & & 9 & & 10 \end{matrix} 
$$ 
The generators of $E_{11}$ to level four [7]  are given by ($\underline{a} = a,11 \ , \ a = 1,..,10$)
$$
\hat{K}^{\underline{a}}{}_{\underline{b}} \ ; \ \hat{R}^{\underline{a}_1 \underline{a}_2 \underline{a}_3} \ ; \ \hat{R}^{\underline{a}_1 .. \underline{a}_6} \ ; \ \hat{R}^{\underline{a}_1 .. \underline{a}_8 , \underline{b}} \ ;  \ 
\hat{R}^{\underline{a}_1 .. \underline{a}_9 , \underline{b}_1 \underline{b}_2 \underline{b}_3} \ , \ $$
\vspace{-0.8cm}
$$
\hat{R}^{\underline{a}_1 .. \underline{a}_{10} , (\underline{b}_1 \underline{b}_2)} \ , \ \hat{R}^{\underline{a}_1 .. \underline{a}_{11} , \underline{b}} \ . \eqno(5.1)$$
along with corresponding negative level generators. The levels are separated by a semi-colon, and all upper indices are anti-symmetric within each block, except for the second level four generator whose last two indices are symmetric, as indicated by the open brackets on those indices. The generators of $E_{11}$ in equation (5.1) are listed as irreducible representations of the $A_{10}$ subalgebra of $E_{11}$, and this is why we say it is an 11D theory.
\par 
The generators of the 10D IIA algebra are given by [7] ($a,b,... = 1,...,10$)
$$
K^a{}_b \ , \ R \ , \ ; R^a \ , \ R^{a_1 a_2} \ , \ R^{a_1 a_2 a_3} \ , \ R^{a_1 ... a_5} \ , $$
$$
R^{a_1 .. a_6} \ , \ R^{a_1 .. a_7} \ , \ R^{a_1 .. a_8} \ , \ ... \eqno(5.2)$$
along with corresponding negative level generators. It is absolutely not obvious that these generators describe a Kac-Moody algebra, let alone the Kac-Moody algebra $E_{11}$.
\par 
We can however embed these generators into an 11D algebra by adding an eleventh index as follows
$$
K^a{}_b = \hat{K}^a{}_b \ , \ R \ , \ ; R^a = 2 \hat{K}^a{}_{11} \ , \ R^{a_1 a_2} = \hat{R}^{a_1 a_2 11}  \ , $$
$$
R^{a_1 a_2 a_3} = \hat{R}^{a_1 a_2 a_3} \ , \ R^{a_1 ... a_5} = - \hat{R}^{a_1 .. a_5 11} \ , $$
$$
R^{a_1 .. a_6} = - \hat{R}^{a_1 .. a_6} \ , \ R^{a_1 .. a_7} = {1 \over 2} R^{a_1 .. a_7 11,11} \ , \eqno(5.3) $$
$$
R^{a_1 .. a_8} = {3 \over 8} \hat{R}^{a_1 .. a_8,11} \ , \ ...$$
It is clear that, on looking at the IIA generators, and simply adding an eleventh index in a minimal manner where possible, we unavoidably end up using the generators of the $E_{11}$ algebra from equation (5.1). In this case, the IIA decomposition of $E_{11}$ is in fact $E_{11}$, just re-written in a different manner, where the eleventh index is suppressed and any generator involving this index is relabelled. Amazingly, a similar situation occurs for the IIB LEEA [3], [7].

\section{6: From $K_{11}$ to $E_{11}$}

Based on the relation between the 10D IIA decomposition of $E_{11}$, and $E_{11}$ describing an 11D theory, we expect that the 10D $K_{11}$ algebra will extend into an 11D algebra by making an 11'th index explicit on the generators in a minimal fashion. The Dynkin diagram of $K_{11} = D_8^{+++}$ is given by 
\[
\begin{matrix}
& & &  & & & \oplus & 10 & & & & & & & \oplus & 11 & & \\
& & &  & & & | & & & & & & & & | & & & \\
\bullet & - & \bullet & - & \bullet & - & \bullet & - & \bullet & - & \bullet & - & \bullet & - & \bullet & - & \bullet \\
1 & & 2 & & 3 & & 4 & & 5 & & 6 & & 7 & & 8 & & 9 \end{matrix} \]
The generators of $K_{11}$, decomposed with respect to $A_9$, are described in terms of integer levels $l = l_{11-1} + l_{11}$ associated to level vectors $\vec{l} = (l_{11-1},l_{11})$, where a generator at level $l$ has $(11-5) l_{11-1} + 2 l_{11}$ indices. The generators of $K_{11}$ to level two are given by ($a,b = 1,...,10$)
$$
K^a{}_b \ (0,0) \ , \ R \ (0,0) \ ; \ R^{a_1 a_2} \ \ (0,1)  \ , \ R^{a_1 .. a_6} \ (1,0) \ ;  $$
$$
R^{a_1 .. a_7,b}  (1,1) ,  R^{a_1 .. a_8}  (1,1) ,  R^{a_1 .. a_9,b_1 b_2 b_3}  (2,0) \ ; \eqno(6.1) $$
Associated to each of these generators are fields $h_a{}^b$, $\phi$, $A_{a_1 a_2}$, $A_{a_1 .. a_{6}}$, $A_{a_1 .. a_{8}}$, $h_{a_1 .. a_{7},b}$ describing the graviton, dilaton, Kalb-Ramond field, dual Kalb-Ramond field, dual dilaton, and dual graviton, respectively.
\par 
Although it is not clear from the Dynkin diagrams, we now argue that $K_{11}$ is a subgroup of $E_{11}$. To do this, we recall the generators of $E_{11}$ to level four from equation (5.2). The $K_{11}$ generators are related to the generators of $E_{11}$ through
$$
K^a{}_b =  \hat{K}^a{}_b \ , \ R = - \sum_{a=1}^{10} \hat{K}^a{}_a + 3 \hat{K}^{11}{}_{11}  \ ; \ $$
\vspace{-0.8cm}
$$
R^{a_1 a_2} = \hat{R}^{a_1 a_2 11} \ , \ R^{a_1 .. a_6} = - \hat{R}^{a_1 .. a_6} \ ;  \eqno(6.2) $$
\vspace{-0.8cm}
$$
 R^{a_1 .. a_7,b} = \hat{R}^{a_1 .. a_7 \ 11 , b} \ , \ R^{a_1 .. a_8} = {3 \over 8} \hat{R}^{a_1 .. a_8,11} \ , $$
\vspace{-0.8cm}
$$
\ R^{a_1 .. a_9,b_1 b_2 b_3}  = \hat{R}^{a_1 .. a_9,b_1 b_2 b_3} \ ; \ ... $$
The $K_{11}$ generators thus occur in the IIA decomposition of $E_{11}$. Since we have described the level one generators of $K_{11}$ in terms of generators obtainable from the level one generators of $E_{11}$, this correspondence will hold at all levels, so that $K_{11}$ must be a subgroup of $E_{11}$. Note, for example, that there is no $R^{a_1 .. a_5} = - \hat{R}^{a_1 .. a_5 11}$ in the $K_{11}$ algebra. Indeed we note that $5 = (11-5) l_{11-1} + 2 l_{11}$ cannot be satisfied in $K_{11}$, so we find that $K_{11}$ is a strict subgroup of (the IIA decomposition of) $E_{11}$.
\par 
We now notice a distinct difference in how the 10D $K_{11}$ algebra is related to the 11D $E_{11}$ algebra, versus how the 10D IIA decomposition of $E_{11}$ is related to the 11D $E_{11}$ algebra. Here $K_{11}$ is a strict subalgebra of $E_{11}$, even when the eleventh index is made explicit. We do not find what we would expect, that $K_{11}$ is a complicated way of re-writing of a simpler 11D algebra, in the manner that the 10D IIA decomposition of $E_{11}$ simplified into $E_{11}$ in 11D. Therefore, directly mimicking how the 10D IIA decomposition of $E_{11}$ extends into an 11D $E_{11}$ algebra, we see the 10D $K_{11}$ just extends into an 11D algebra of the same form with an inert eleventh index carried around, and it can be embedded in $E_{11}$ as a strict subalgebra, we do not get anything differing non-trivially from $K_{11}$ as it appears in 10D.
\par 
The only obvious non-trivial way to extend $K_{11}$ occurs when we try to describe a theory in 11D in the way that `bosonic m-theory' did [10], by utilizing the full 3-form in 11D. Extending $R^{a_1 a_2} = \hat{R}^{a_1 a_2 11}$ into the full $E_{11}$ generator $\hat{R}^{\underline{a}_1 \underline{a}_2 \underline{a}_3}$, noting that gravity comes along with $E_{11}$ at level zero automatically, this implies that we immediately find all of $E_{11}$, as $E_{11}$ is generated by the level one generator. There can thus be no strict algebra between $K_{11}$ and $E_{11}$ describing gravity and a 3-form, it has to be $E_{11}$. This immediately implies that `bosonic m-theory' is either trivially just $K_{11}$ with an inert eleventh index, or non-trivially the usual (bosonic sector of) m-theory based on $E_{11} \otimes_s l_1 /I_c(E_{11})$. 
\par 
In other words, a non-trivial `11D bosonic m-theory' containing gravity and a 3-form is (the bosonic sector of) m-theory in 11D. It is simply unavoidable that the minimal 11D theory involving gravity and a 3-form which contains $K_{11}$ must be $E_{11}$, and so describe the usual (bosonic sector of) m-theory. Without supersymmetry arguments, $E_{11}$ offers the only means to directly obtain the 11D bosonic SUGRA action [1, 2, 5], it contains $K_{11}$, and $K_{11}$ offers a direct way to obtain the 10D CBS with the correct coefficients. A consistent theory based on $K_{11}$ thus extends to a consistent theory based on $E_{11}$ in 10D $\to$ 11D. 

\section{7: $K_{27}$ is Not a Subalgebra of $E_{27}$}

We now consider the 26D $\to$ 27D transition. Starting from $K_{27}$, and mimicking what we have done in the previous section, going from the 26D CBS to 27D `bosonic m-theory' should amount to the claim that $K_{27}$ is a subalgebra of $E_{27}$, that a dynamical theory of duality relations in 27 dimensions based on $E_{27}$ is consistent, and that it contains the consistent dynamical theory of the 26D CBS based on $K_{27}$ [9]. Indeed, given that the `bosonic m-theory' of [10] is based on gravity and a 3-form, it simply has to be the case that the presumptive Kac-Moody symmetry of `bosonic m-theory' would be $E_{27}$, where level zero describes gravity, and the level one generator of $E_{27}$ is a 3-form. However we will now argue $K_{27}$ is not a subalgebra of $E_{27}$, and in the following section we argue that a theory of duality relations based on $E_{27}$ can not exist. Thus, while a consistent theory of 26D closed strings based on $K_{27}$ does exist, no obvious `bosonic m-theory' extension is possible in 27D, and it would have to unnaturally diverge from the manner in which $K_{11}$ extends to $E_{11}$, a possibility we will not consider.
\par 
The Dynkin diagram of $K_{27} = D_{24}^{+++}$ is given by
$$
\begin{matrix}
& & &  & & & \oplus & 26 & & & & & & & \oplus & 27 & & \\
& & &  & & & | & & & & & & & & | & & & \\
\bullet & - & \bullet & - & \bullet & - & \bullet & - & \bullet & - & \cdots & - & \bullet & - & \bullet & - & \bullet \\
1 & & 2 & & 3 & & 4 & & 5 & & & & 23 & & 24 & & 25 \end{matrix} 
$$ 
The generators of $K_{27}$ decomposed with respect to $A_{25}$ are described in terms of levels $l = l_{27-1} + l_{27}$ associated to level vectors $\vec{l} = (l_{27-1},l_{27})$, where a generator at level $l$ has $(27-5) l_{27-1} + 2 l_{27}$ indices. The generators of $K_{27}$, partially to level two, are given by [9] $(c,d = 1,...,26$)
$$
K^c{}_d  \ \ (0,0) \ \ , \ \ R \ \ (0,0) \ \ ; \ \ R^{c_1 c_2} \ \ (0,1) \ \ , \ \ R^{c_1 ... c_{22}} \ \ (1,0) \ \ ; $$
\vspace{-0.8cm}
$$
R^{c_1 ... c_{24}} \ \ (1,1) \ \ , \ \ R^{c_1 .. c_{23},d} \ \ (1,1) \ \ . \eqno(7.1)  $$
Associated to each of these generators are fields $h_a{}^b$, $\phi$, $A_{c_1 c_2}$, $A_{c_1 .. c_{22}}$, $A_{c_1 .. c_{24}}$, $h_{c_1 .. c_{23},d}$ describing the graviton, dilaton, Kalb-Ramond field, dual Kalb-Ramond field, dual dilaton, and dual graviton, respectively.
\par 
The Dynkin diagram of $E_{27}$, decomposed with respect to its $A_{26}$ subgroup, is given by
$$
\begin{matrix}
& & &  & & & & & & & \oplus & 27 & & \\
& & &  & & &  & & & & | & & & \\
\bullet & - & \bullet & - & \bullet & - & ... & - & \bullet & - & \bullet & - & \bullet & - & \bullet \\
1 & & 2 & & 3 & & & & 23 & & 24 & & 25 & & 26 \end{matrix} 
$$ 
The generators of $E_{11}$ are a subset of the generators of $E_{27}$, thus the generators of equation (5.1) appear in both $E_{11}$ and $E_{27}$. The generators of $E_{11}$ to level 8 are listed in [7]. Beyond level four, one finds additional generators in $E_{27}$. These generators can be listed at low levels via the computer program [12], however the computational difficulty means it was not possible to fully list $E_{27}$ to level 8, but we believe it contains generators such as an $R^{23,1}$ and an $R^{22,(1,1)}$ ($\underline{c}=c,27$)
$$
\hat{K}^{\underline{c}}{}_{\underline{d}} \ ; \ \hat{R}^{\underline{c}_1 \underline{c}_2 \underline{c}_3} \ ; \ ... \ ; \ \ \hat{R}^{\underline{c}_1 .. \underline{c}_{23},\underline{d}} \ , \ \hat{R}^{\underline{c}_1 .. \underline{c}_{22},(\underline{d}_1 \underline{d}_2)} \ . \eqno(7.2)$$
We associate to these generators the fields $h_{\underline{c}}{}^{\underline{d}}$, describing the graviton, and $A_{\underline{c}_1 \underline{c}_2 \underline{c}_3}$, which is precisely the three-form conjectured in [10] to form part of `bosonic m-theory'. 
\par 
The correspondence between $K_{27}$ and $E_{27}$ would now be something of the form
$$
K^c{}_d = \hat{K}^c{}_d \ , \ R = - \sum_{c=1}^{26} \hat{K}^c{}_c + 3 \hat{K}^{27}{}_{27} \ ; \ $$
\vspace{-0.5cm}
$$
R^{c_1 c_2} = \hat{R}^{c_1 c_2 27} \ , \ R^{c_1 .. c_{22}} = \hat{R}^{c_1 .. c_{22} 27 , 27} \ ; \eqno(7.3) $$
In order for a correspondence like this to be valid, we would need the $K_{27}$ relation $[R^{c_1 c_2},R_{d_1 .. d_{22}}] = 0$ (equation (2.7) of [9]) to hold in terms of the underlying $E_{27}$ generators, namely we need 
$$
[\hat{R}^{c_1 c_2 27},\hat{R}_{d_1 .. d_{22} 27,27}] =^? 0. \eqno(7.4)$$
If equation (7.4) is non-zero in $E_{27}$, then $K_{27}$ cannot be a subalgebra of $E_{27}$. Due to the extreme difficulty, computational and practical, we do not have the algebra of $E_{27}$ to level 8, so we cannot check this explicitly. However, we can be sure that this commutator is non-zero because $R^{\underline{c}_1 \underline{c}_2 \underline{c}_3}$ is the level one generator of $E_{27}$, and so the right-hand side of equation (7.4) would be the most general sum of generators at level seven in $E_{27}$ with the correct index structure. The only way it can vanish is because the choice of the indices as ``27" ensures it always vanishes, however it does not always need to vanish. As an example, consider the $E_{11}$ level four commutator $[R^{a_1 a_2 11},R_{b_1 .. b_{10} 11,11}] = \frac{495}{4} \delta^{a_1 a_2 11}_{[b_1 b_2 11} R_{b_3 .. b_{10}],11}$. The ``[level 1,level minus 8] = level minus 7" analog would certainly have many non-zero terms on the right-hand side of the analogous relation. 
\par 
Conceptually, looking at the level one $K_{11}$ generators in equation (6.2), the 10D Kalb-Ramond field comes from the 11D 3-form, and the dual Kalb-Ramond field comes from the dual of the 3-form, which makes physical sense. For $K_{27}$ and $E_{27}$, the Kalb-Ramond field of $K_{27}$ comes from the $E_{27}$ 3-form, however the dual of the Kalb-Ramond field in equation (7.3) would now be coming from an $R^{23,1}$ that has the form of a generator which usually describes a dual-graviton. This does not make physical sense and it does not match the 10-D $\to$ 11-D logic.
\par 
One may argue that the $R^{c_1 .. c_{22}}$ of $K_{27}$ simply needs to be associated to some other generator of $E_{27}$, for example that we should set $R^{c_1 .. c_{22}} = \hat{R}^{c_1 .. c_{22},(27 \ 27)}$. However the same non-vanishing commutator problem arises, comparing to the analogous non-vanishing $E_{11}$ ``[level 1, level minus 4] = level minus 3" commutator $[\hat{R}^{a_1 a_2 11},\hat{R}_{b_1 .. b_{10},(11 \ 11)}]$. Physically, it is very likely that the $E_{27}$ field $\hat{R}^{\underline{c}_1 .. \underline{c}_{22},(\underline{d}_1 \underline{d}_2)}$ carries no degrees of freedom, in the same manner that the $E_{11}$ level four field $\hat{R}^{\underline{a}_1 .. \underline{a}_{10},(\underline{b}_1 \underline{b}_2)}$ carries no degrees of freedom [13], so a 27D field with no degrees of freedom containing the 26D dual Kalb-Ramond makes little physical sense. 
\par 
The problem from a physical perspective, as we shall see in the next section explicitly, is that there is e.g. no dual of the three-form in $E_{27}$, so there is no natural way to associate the 26D dual Kalb-Ramond field $R^{c_1 .. c_{22}}$ to the dual of the 27D 3-form, hence a consistent theory of duality relations based on $E_{27}$ cannot exist.
Thus, the `27D bosonic m-theory algebra' should be $E_{27}$, and it should contain both the $(0,1)$ and $(1,0)$ generators of $K_{27}$, however $E_{27}$ does not contain a dual three-form `parent' for the $(1,0)$ generator.

\section{8: `Bosonic M-Theory' Does Not Seem to Exist in 27D}

In order for a consistent `bosonic m-theory' to exist based on $E_{27}$, we must be able to describe the dynamics of the 27D graviton $h_{\underline{a}}{}^{\underline{b}}$, and the three form $A_{\underline{a}_1 \underline{a}_2 \underline{a}_3}$, in terms of first order duality relations involving fields at higher levels in $E_{27}$, where the covariant derivatives used in the duality relations arise from the Maurer-Cartan form $g^{-1} d g$ in the non-linear realization $E_{27} \otimes_s l_1 / I_c(E_{27})$. The duality relation involving the three-form must read as [2, 5]
$$
D_{\underline{c}_1 .. \underline{c}_4} = \hat{G}_{[\underline{c}_1,\underline{c}_2 \underline{c}_3 \underline{c}_4]} + e_1 \varepsilon_{\underline{c}_1  .. \underline{c}_4}{}^{\underline{d}_1 .. \underline{d}_{23}} \hat{G}_{\underline{d}_1,\underline{d}_2 .. \underline{d}_{23}} \eqno(8.1)$$
for $e_1$ (and $e_2$ below) a coefficient that would be uniquely determined by the symmetries of the non-linear realization. We thus need an $A_{\underline{c}_1 .. \underline{c}_{22}}$, but there is no $A_{\underline{c}_1 .. \underline{c}_{22}}$ in $E_{27}$. There is thus no consistent theory of first order duality relations relating the fields of $E_{27}$ to one another. Contrast this with the situation in $K_{27}$ in 26 dimensions [9]. In $K_{27}$, one indeed finds an $A_{a_1 a_2}$ and its dual $A_{a_1 .. a_{22}}$, related by the duality relation (4.0.2) of [9] which is of a similar form to that of equation (8.1).
\par 
Similarly, a duality relation between the graviton and a prospective dual graviton must take the form [2,5]
\begin{small}
$$ 
D_{\underline{c},\underline{d}_1 \underline{d}_2} = (\det e)^{{1 \over 2}} \hat{\omega}_{\underline{c},\underline{d}_1 \underline{d}_2} + e_2 \varepsilon_{\underline{d}_1 \underline{d}_2}{}^{\underline{e}_1 .. \underline{e}_{25}} \hat{G}_{\underline{e}_1,\underline{e}_2 .. \underline{e}_{25},\underline{c}} \eqno(8.2)$$ \end{small}
In order for this duality relation to be consistent we need an $\hat{R}^{\underline{c}_1 .. \underline{c}_{24},\underline{d}}$ in $E_{27}$ to find a dual graviton $\hat{h}_{\underline{c}_1 .. \underline{c}_{24},\underline{d}}$, but there is no such generator. In $E_{27}$ the closest generator is an $\hat{R}^{23,1}$, however this is associated to the dual graviton in $K_{27}$ over ${\rm GL}(26)$, but $E_{27}$ decomposes with respect to ${\rm GL}(27)$. This illustrates the very special nature of $E_{11}$, only in $D=11$ will $E_{D}$ lead to duality relations involving the 3-form and it's dual, and the graviton linked to a dual-graviton, without additional dynamical fields with no interpretation. 
\par 
We now comment on an attempt to modify $E_{27}$ to force, for example, a dual 3-form into the theory. Trying to shoehorn in a dual of the 3-form, by adding additional nodes to $E_{27}$ to create a new Dynkin diagram, produces a proliferation of additional fields, forces us to add a dilaton in the initial 27D theory, all in a way that radically diverges from the pattern set between $K_{11}$ and $E_{11}$, and so we do not consider such a modification further. Thus, a non-trivial `bosonic m-theory', as motivated by the analogy of going from the 10D IIA string to 11D m-theory, when restricted to the CBS, does exist in 10D $\to$ 11D, where it is the (bosonic sector of) m-theory, and does not seem to exist in 26D $\to$ 27D.

\section{9: Summary and Future Work}

We have analyzed, in terms of the Kac-Moody algebras $K_{11}$ and $E_{11}$, the idea of a non-trivial 11D `bosonic m-theory' extension of the 10D CBS, and argued that it must be (the usual bosonic sector of) m-theory. This was immediately forced on us when we extended $\hat{R}^{a_1 a_2 11}$ into the full level one $E_{11}$ generator $\hat{R}^{\underline{a}_1 \underline{a}_2 \underline{a}_3}$. We then argued that in 26D, the analogous transition to `bosonic m-theory' would involve going from $K_{27}$ to $E_{27}$, noting $E_{27}$ contains the precise three-form required for the presumptive `bosonic m-theory'. However the level $(1,0)$ dual-Kalb-Ramond generator $\hat{R}^{22}$ of $K_{27}$ has no natural dual `parent' in $E_{27}$. We claimed that $K_{27}$ is not a subalgebra of $E_{27}$, that a theory of dualities based on $E_{27}$ does not exist, and so that a non-trivial 27D `bosonic m-theory' which contains $K_{27}$ does not obviously exist. Containing $K_{11}$ and $K_{27}$ is a very useful requirement when seeking a `bosonic m-theory', as this enables one to find related algebras in a direct manner [14]. It would be interesting to relate this work to the so-called `monstrous m-theory' in 26+1 dimensions of [15], especially in light of the fact that `bosonic m-theory' was used as evidence in favor of `monstrous m-theory', and in light of the propsed link of $K_{27}$ to monstrous moonshine discussed in [1].

\section{Acknowledgements}

We would like to thank Peter West, Paul Cook, Mirian Tsulaia, Yasha Neiman, and David Chester, for discussions, particularly Paul Cook for help with $E_{27}$ at level 8. This work was supported by the Quantum Gravity Unit of the Okinawa Institute of Science and Technology Graduate University (OIST).

\end{document}